%% file: McDaid_COMPSTAT2012.tex
\documentclass[11pt,a4paper,oldfontcommands]{proceedings_CFE}
\usepackage[numbers,sectionbib]{natbib}
\usepackage{hyperref}
\usepackage{url}
\begin{document}

\mainmatter
\pagestyle{proceedings}

%%%%%%%%%%%%%%%%%%%%%%%%%%%%%%%%%%%%%%%%%%%%%%%%%%%%%%%%%%%%%%%%%%%%%%%%%%%%%%%%%%%%%%%%%%%%%%%%%%%%%%%%%%%%%%%%%%%%%%%%
%%%%%%%%%%%%%%%%%%%%%%%%%%%%% Here comes your COMPSTAT 2012 manuscript %%%%%%%%%%%%%%%%%%%%%%%%%%%%%%%%%%%%%%%%%%%%%%%%%
%%%%%%%%%%%%%%%%%%%%%%%%%%%%%%%%%%%%%%%%%%%%%%%%%%%%%%%%%%%%%%%%%%%%%%%%%%%%%%%%%%%%%%%%%%%%%%%%%%%%%%%%%%%%%%%%%%%%%%%%

% ----------------------------------------------------------------------------------------------------------------------
	% \COMPSTATManuscript{Authors}{Long title (at most 15 words)}{Short title (for header)}

% Examples: (please modify the corresponding data as you wish)
% Single Author
%\COMPSTATManuscript{John Smith}{This is the long manuscript title: at most 15 words}{Short manuscript title (for header)}

% Multiple authors
%\COMPSTATManuscript{John Smith \textit{et al.}}{This is the long manuscript title: at most 15 words}{Short manuscript title (for header)}
% or (if enough room)
\COMPSTATManuscript{Aaron F. McDaid, Thomas Brendan Murphy, Nial Friel and Neil Hurley}{Model-based clustering in networks with Stochastic Community Finding}{Stochastic Community Finding}
%-----------------------------------------------------------------------------------------------------------------------

%-----------------------------------------------------------------------------------------------------------------------
% Declaration of the different authors (please modify the corresponding data as you wish)
%\COMPSTATAuthor{Full Author Name}{Institution}{Email}

% Example: (use a line for each author)
\COMPSTATAuthor{Aaron F. McDaid}{University College Dublin, Ireland}{aaronmcdaid@gmail.com} %Add a line for each author
\COMPSTATAuthor{Thomas Brendan Murphy}{University College Dublin, Ireland}{brendan.murphy@ucd.ie} %Add a line for each author
\COMPSTATAuthor{Nial Friel}{University College Dublin, Ireland}{nial.friel@ucd.ie} %Add a line for each author
\COMPSTATAuthor{Neil J. Hurley}{University College Dublin, Ireland}{neil.hurley@ucd.ie} %Add a line for each author
%-----------------------------------------------------------------------------------------------------------------------

%-----------------------------------------------------------------------------------------------------------------------
% Declaration of the abstract
%\COMPSTATAbstract{Abstract paragraph}

% Example: (please modify the corresponding data as you wish)
\COMPSTATAbstract{
In the model-based clustering of networks, \emph{blockmodelling} may be used to
identify roles in the network.
We identify a special case of the Stochastic Block Model (SBM) where
we constrain the cluster-cluster interactions such that
the density \emph{inside} the clusters of nodes is expected to be greater than the density
\emph{between} clusters.
This corresponds to the intuition behind \emph{community-finding} methods,
where nodes tend to clustered together if they link to each other.
We call this model Stochastic Community Finding (SCF) and present an efficient
MCMC algorithm which can cluster the nodes, given the network. % and select the number of clusters.
The algorithm is evaluated on synthetic data and is applied to a social network
of interactions at a karate club and at a monastery,
demonstrating how the SCF finds the `ground truth' clustering where sometimes the SBM does not.
The SCF is only one possible form of constraint or specialization that may
be applied to the SBM.
In a more supervised context, it may be appropriate to use other specializations
to guide the SBM.

% The use of \emph{collapsing} and modern algorithmic techniques such as the
% \emph{allocation sampler} allow us to develop this MCMC algorithm
% and avoid the complications of RJMCMC.
% We are able to directly estimate the number of clusters in the algorithm,
% avoiding the need for a separate model selection criterion such as the BIC or ICL.

\iffalse
Each manuscript must be preceded by an abstract of
  200 words at most in a single paragraph that summarizes the
  content. As an illustration this one contains about 100 words.
  Please do not capitalize letters. Titles and names should be written
  with lower letters except from the first letter which should be in
  capitals. E.g. The estimation of GARCH models. John Smith.  Simple
  Latex syntax can be used, e.g. $\hat{x}_i$, $\beta$, etc. The
  abstract should not list or contain any references (e.g. do not have
  statements like "... in John Smith (2011)").  {\em The authors\/},
  etc. should be avoided.  The title of the manuscript should be at
  most 15 words.
  \fi
  }
%-----------------------------------------------------------------------------------------------------------------------

%-----------------------------------------------------------------------------------------------------------------------
% Declaration of the keywords
%\COMPSTATKeywords{Keyword1, Keyword2, etc.}

% Example: (please modify the corresponding data as you wish)
\COMPSTATKeywords{Model-based clustering, MCMC, Social networks, Community finding, Blockmodelling}
%-----------------------------------------------------------------------------------------------------------------------

\input{intro}
\input{sbm}
\input{scf}
\input{collapse}
\input{algo}
\input{synth}
\input{karate}

\input{conclusion}

\subsection*{Acknowledgement}
This research was supported by Science Foundation Ireland (SFI) Grant No. 08/SRC/I1407.

%-----------------------------------------------------------------------------------------------------------------------
\bibliographystyle{year}
\bibliography{aaron}
%-----------------------------------------------------------------------------------------------------------------------

% TODO:
% \begin{itemize}
	% \item Related Work section?
% \end{itemize}

\end{document}

%% file: intro.tex
\section{Introduction}
\label{SECintro}

Clustering typically involves dividing objects into clusters where
objects are in some sense `close~to' the other objects in the same cluster.
Much research has been done into clustering points in Euclidean space
where points are put into the same cluster based on a distance metric between pairs points.
But the data we have is of a different form, we have a network as input data.

In network analysis, clustering is usually based on the idea that two nodes in the network are `close to' each other if
they are linked to each other. This is called \emph{community-finding} and is the main topic of this paper.
% This broad idea has appeared often in the social science literature.
There are a large number of methods using heuristic algorithms and non-statistical objective functions \cite{NewmanGirvan,rosvall-2008,BansalCorrelationClustering}.
% There are also algorithms which allow the communities to overlap with each other\cite{LancichinettiOSLOM,palla-2005}.
The complexity issues around some such algorithms are also discussed in the literature \cite{CharikarQuadratic, DasGuptaComplexityOfModularity}.
For a thorough review of the broad area of research into clustering the nodes of a network, see \cite{fortunato-2010}.

In the rest of this paper we focus on statistical models and algorithms, as they are relevant for our approach.
We base our model in the Stochastic Block Model (SBM) of \cite{Nowicki-01}.  That model is not, by default,
a community-finding model.  For example, with the famous social network known as Zachary's Karate Club
the SBM will, if asked for two clusters, divide the nodes into one small cluster of high-degree nodes
and another cluster containing a large number of smaller-degree nodes.
In community-finding, this would be seen as an `incorrect' result;
the members of the karate club went on to divide themselves into two factions, where most of the friendship
edges are, unsurprisingly, inside the factions.
Community-finding methods are expected to find this type of clustering, where the edges tend to be inside clusters.

Many of the probabilistic models of networks are based on the SBM \cite{Daudin-08, ZanghiOnline}
and therefore they do not explicitly tackle community-finding.
In this paper, we make a change to the standard SBM to require that the blocks corresponding to
within-cluster connectivity will be expected to be denser than the blocks corresponding to
between-cluster connectivity.
This will lead to an algorithm which, unlike the SBM, will cluster the nodes according to the two factions in the karate club,
as would be expected in a community-finding algorithm.

Given a generative model and an observed network, we can check the posterior distribution and
obtain a clustering, or set of clusterings, which are a good fit for the data.
It is typically trivial to write MCMC algorithms to sample from the relevant distribution.
However, it can be challenging to create suitably fast algorithms.
We use \emph{collapsing} along with algorithmic techniques such as the \emph{allocation sampler}
\cite{NobileAllocationSampler}; a scalable application of these ideas to the standard SBM is in \cite{McDaidSBM}.
% The question of scalability arises in any algorithm.  Much of the computer science literature looks
% Traditionally, MCMC methods have been perceived as begin slow.
% However, MCMC performance is very dependent on the proposal heuristic that is used and recent research in MCMC
% algorithms \cite{NobileAllocationSampler} suggests that there is no limit to how efficient and scalable an MCMC method can be.
% If the heuristic correlates well with the desired distribution, then the acceptance rates and mixing of the Markov Chain will be extremely good.
% We expect further progress will be made in the literature.

% [ reference the Smola paper on collapsing? ]

In applying these concepts to the SCF we run into a problem though. It does not appear to be possible
to directly integrate out the relevant parameters to give us a fully collapsed model.
However, we will show in this paper how
we can work around this and still develop a suitable Metropolis-Hastings algorithm with the correct
transition probabilities without having to resort to trans-dimensional RJMCMC\cite{GreenRJMCMC}.
This technique is not a typical application of Metropolis-Hastings and it may have broader applicability,
allowing faster algorithms with the simplicity of collapsing, in models where full explicit collapsing is not possible.

\subsection{Structure of this paper}
In Section \ref{SECsbm} we will review the standard SBM of \cite{Nowicki-01} - defining the
basic notation and models which will be used throughout.
In Section \ref{SECscf} we will define our modification to the SBM which we call Stochastic Community Finding (SCF).
% [ \ldots related work section? \ldots ]
In Section \ref{SECcollapse} we will consider the issue of collapsing; this is straightforward for the SBM,
but not for the SCF.
In Section \ref{SECalgo} we discuss the algorithm used in our
software\footnote{C++ implementation, and datasets used, at \url{https://sites.google.com/site/aaronmcdaid/sbm}}
which enables us to use Metropolis-Hastings even though we cannot write down the collapsed posterior mass in closed form.
We then proceed to evaluations, first considering a synthetic network in Section~\ref{SECsynth} and finally
an analysis of Zachary's Karate Club and Sampson's Monks in Section~\ref{SECkarate}.
We close with a discussion of possible future directions in Section~\ref{SECconclusion}.

%% file: sbm.tex
\section{Stochastic Block Model}
\label{SECsbm}
In this section, we define the Stochastic Block Model (SBM) of \cite{Nowicki-01} before discussing our modification in the next section.
We restrict our attention in this section to directed unweighted networks, where edges are simply present or absent.
There are many extensions\footnote{ directed \emph{or} undirected, unweighted \emph{or} integer-weights and other more complex `alphabets' to describe an edge, self-loops modelled \emph{or} ignored.},
for example allowing weighted networks with integer- or real-valued weights \cite{McDaidSBM, WyseFriel}.
% Undirected networks can be naturally handled also.
% Our SCF modification can be applied to all of them, but in this evaluation we will focus on the simpler directed unweighted model.

We model a network of $N$ nodes, and the network is represented as an adjacency matrix $x$.
If there is a directed edge from node $i$ to $j$, we have $x_{ij}=1$.
If they are not connected, we have $x_{ij}=0$.
By default, we ignore self loops ($x_{ii}$) and they are simply left out of the formulae.

Given a network $x$, our goal is to identify a clustering $z$. We use a vector $z$ of length $N$,
where $z_i$ is the cluster to which node $i$ is assigned. There are $K$ clusters, $1 \leq z_i \leq K$.

Given $K$ clusters, there are $K\times K$ \emph{blocks}, one block for each pair of clusters.
There is a $K \times K$ matrix $\pi$ which records, for each block, the expected density of
edge-formation in that block.
In other words, given node $i$ which is in cluster $k = z_i$ and node $j$ which is in cluster $l = z_j$,
the probability of a connection is $\pi_{kl}$,

$$ x_{ij} \sim \text{Bernoulli}(\pi_{kl}) $$

In the undirected variant we would have $x_{ij}=x_{ji}$, and only a single draw from the relevant
Bernoulli would be used to assign to these.
The probability of two nodes connecting depends on the clusters to which the nodes are assigned, but
is otherwise independent of the particular nodes; this is the definition of blockmodelling.
The elements of $\pi$ have a prior; $\pi_{kl} \sim \text{Beta}(\beta_1,\beta_2)$.
Our default is to set $\beta_1=\beta_2=1$ which means this prior is a Uniform
distribution over (0,1).

$z$ is itself a random variable. There is a vector $\theta$ of length $K$ which represents the
probability, for each cluster, of any node being assigned to that cluster.
$ z_i \overset{iid}\sim \text{Multinomial}(1; \theta_1, \theta_2, \cdots \theta_K) $
% $\theta$ is constrained such that $1 = \sum_{k=1}^K \theta_k$.

$\theta$ is also a random variable and we place a Dirichlet prior on it.

\begin{equation}
	\theta \sim \text{Dirichlet}(\alpha_1, \alpha_2, \cdots, \alpha_K)
	\label{EQdirichlet}
\end{equation}

The parameters to the Dirichlet prior are a choice to be made by the user,
and it is conventional to set each of the $\alpha_k$ to the same value, $\alpha_k = \alpha$, and
we set $\alpha$ to 1 by default in our experiments.

Given $N$ and $K$, this is a fully specified generative model to generate many variables including
the clustering $z$ and the network $x$.
We investigated this model in \cite{McDaidSBM}. An important extension we introduced there is to place a prior
on $K \sim \text{Poisson}(1) $,
thus allowing us to deal directly with the number of clusters as a random variable and avoids the
need for any separate model selection criterion. See that paper for a more extended discussion of
model selection and validation of the accuracy of the method in estimating the number
of clusters. %, $K$, and finding the clustering, $z$.

% $N$ is not a variable. We simply write

$$ \mathrm{P}(x,\pi,z,\theta,K) = \mathrm{P}(K) \times \mathrm{p}(z,\theta|K) \times \mathrm{p}(x,\pi|z,K) $$

where we use $\mathrm{P}(\dots)$ for probability mass functions, i.e. of discrete quantities such as $z$ or $K$,
and $\mathrm{p}(\dots)$ for probability density functions.

%% file: scf.tex
\section{Stochastic Community Finding}
\label{SECscf}
Now that we have defined the SBM, as introduced by \cite{Nowicki-01}, % and extended by \cite{McDaidSBM},
we define the modification we are introducing in the Stochastic Community Finding (SCF) model.
In community-finding, as opposed to block-modelling, we expect that if a pair of nodes are
connected then the nodes are more likely to be clustered together than if they were not connected.

$$ \mathrm{P}(z_i = z_j | x_{ij}=1) > \mathrm{P}(z_i = z_j | x_{ij}=0) $$

Blockmodelling doesn't have such a constraint.
This is not a hard rule in community-finding, it is a useful guide to help define the different
goals in community-finding and block-modelling.
An equivalent statement is

$$ \mathrm{P}( x_{ij}=1 | z_i = z_j) > \mathrm{P}(x_{ij}=1 | z_i \neq z_j) $$

This is the formulation we use to define the SCF.
We require that all the diagonal entries in $\pi$ be larger than the off-diagonal entries of $\pi$.
$ \min(\pi_{mm}) > \max(\pi_{kl})$ for all $m,k,l$ where $k \neq l$.
% In terms of the generative model, you can consider that $K$ is generated first as
% per its prior and then the $K \times K$ matrix $\pi$ is drawn with the Beta prior
% for each element.
% If $\pi$ does not satisfy the constraint that the diagonal elements are larger than the
% off-diagonal elements, then it is abandoned \emph{along with $K$}.
% The generative process involves (re-)generating $K$ and $\pi$ until the constraint is
% satisfied.
Define a function $v(\pi)$ which returns 1 if $\pi$ satisfies the constraint, and
returns 0 if it does not.

\begin{equation}
v(\pi) = \left\{ \begin{array}{ccl}
	1 & & \text{if} \; \min(\pi_{mm}) > \max(\pi_{kl}) \qquad \text{for} \; k \neq l \\
	0 & & \text{otherwise}

\end{array} \right.
\label{EQv}
\end{equation}

Under this constraint, the probability density of the SCF model is proportional to $f(x,\pi,z,\theta,K)$
where

$$ f(x,\pi,z,\theta,K) = \mathrm{P}_\text{SBM}(x,\pi,z,\theta,K) \times v(\pi) $$

and $ \mathrm{P}_\text{SBM} $ is the probability density as defined by the SBM.
This probability mass function is essentially identical to the SBM except that we have set
the density to zero where the constraint on $\pi$ is not satisfied.
%$f$ is not a probability distribution as it does not sum to 1 if summed/integrated over all the values
%of its parameters. But this is not a problem for MCMC methods; given any non-negative function
%it is straightforward to use it in a Metropolis-Hastings algorithm will will draw from the
%distribution which is \emph{proportional} to the factor $f$.
%$$ \mathrm{P}_\text{SCF}(x,\pi,z,\theta,K) = \frac{ f(x,\pi,z,\theta,K) }{ \sum_{K,z,x} \int f (x,\pi,z,\theta,K) \mathrm{d}\pi \mathrm{d}\theta } $$
%This latter formula is of no further use to us. We define $f$ by equality and in due course we will
%describe an algorithm to sample from the distribution proportional to $f$.
A simpler form of the SBM has been investigated \cite{ZanghiOnline} where all the diagonal entries
in the blockmodel are taken to be equal to $\lambda$ and the all the off-diagonal
entries are equal to $\epsilon$.
Their model does not explicitly require that $\lambda > \epsilon$, and
hence it is not quite a community-finding model.

%% file: collapse.tex
\section{Collapsing}
\label{SECcollapse}
Given a network $x$, our goal is to estimate the number of clusters and to find the clustering $(K,z)$.
In the SBM as investigated by \cite{McDaidSBM}, it is straightforward to use \emph{collapsing}
and integrate out the other variables that we are not directly interested in such as 
$\pi$ and $\theta$,

\begin{align*}
	\mathrm{P}_\text{SBM}(x,z,K) & = \mathrm{P}_\text{SBM}(K)   \times   \mathrm{P}_\text{SBM}(z|K)                                \times   \mathrm{P}_\text{SBM}(x|z,K) \\
	                             & = \mathrm{P}_\text{SBM}(K)   \times   \int \mathrm{P}_\text{SBM}(z,\theta|K) \;\mathrm{d}\theta   \times   \int \mathrm{P}_\text{SBM}(x,\pi|z,K) \;\mathrm{d}\pi 
\end{align*}
allowing one to create an algorithm which, given $x$, samples $(z,K)$.
% from a distribution which is proportional to $\mathrm{P}_\text{SBM}(x,z,K)$ where
% $x$ is the observed network; this gives us the posterior distribution $z,K | x$.

But this collapsing does not work in such a straightforward way with the SCF;
we cannot, to our knowledge, write down a closed form expression for $f(x,z,K)$ where $\pi$ and $\theta$ have been integrated out.
The problem is that it is difficult to integrate out $\pi$ in the SCF due to the dependence structure between the blocks which is
introduced by the constraint in Equation \ref{EQv}.
In the SBM, the elements of $\pi$ are independent of each other.
Also, given $z$, the various blocks within $x$ which correspond to the elements of $\pi$ are independent of each other and dependent only on a single element of $\pi$.

The model for $K$ and $\theta$ and $z$ are the same in the SCF as in the SBM, therefore we will simply
use $\mathrm{P}(\dots)$ and $\mathrm{p}(\dots)$ for these. But for expressions involving $\pi$ it will make sense to use $\mathrm{p}_\text{SBM}(\dots)$
and $f(\dots)$ to distinguish between the (normalized) probability distribution of the SBM and
the (non-normalized) function for the SCF.
We attempt to collapse as much as possible in order to get an expression for $f(x,z,K)$, our desired stationary distribution:

%\begin{equation*}
	%f(x,z,K) = \mathrm{P}(K) \times      \mathrm{P}(z       |K)                    \times \int \mathrm{p}_\text{SBM}(x,\pi |z,K) \times v(\pi) \;\mathrm{d}\pi
%\end{equation*}

\begin{equation}
	\begin{split}
	f(x,z,K) & = \mathrm{P}(K) \times \mathrm{P}(z|K) \times \int \mathrm{p}_\text{SBM}(x,\pi |z,K) \times v(\pi) \;\mathrm{d}\pi \\
	         & = \mathrm{P}(K) \times \mathrm{P}(z|K) \times \int \mathrm{P}_\text{SBM}(x|z,K) \times \mathrm{p}_\text{SBM}(\pi | x,z,K) \times v(\pi) \;\mathrm{d}\pi \\
	         & = \mathrm{P}(K) \times \mathrm{P}(z|K) \times \mathrm{P}_\text{SBM}(x|z,K) \times \int \mathrm{p}_\text{SBM}(\pi | x,z,K) \times v(\pi) \;\mathrm{d}\pi \\
	         % & = \mathrm{P}(K) \times \mathrm{P}(z|K) \times \mathrm{P}_\text{SBM}(x|z,K) \times \mathrm{P}_\text{SBM}(v(\pi)=1 | x,z,K) \\
	         & = \mathrm{P}_\text{SBM}(x,z,K) \times \mathrm{P}_\text{SBM}(v(\pi)=1 | x,z,K)
	\end{split}
\label{EQfFull}
\end{equation}

The final factor in the final expression, $\mathrm{P}_\text{SBM}(v(\pi)=1 | x,z,K)$, can be interpreted as the probability (under the SBM),
given $(x,z,K$),
that a draw of $\pi$ will satisfy the constraint; it is this factor that, to our knowledge, cannot be solved in closed form.
The first factor in the final expression, $ \mathrm{P}_\text{SBM}(x,z,K) $, can be directly taken from \cite{McDaidSBM}
as the relevant integration has been solved as described in the Appendices of that paper.
In the following expression, we define $n_k$ to be the number of nodes in cluster $k$, i.e. $n_k$ is a function of $z$.
Also, $p_{kl}$ is the number of pairs of nodes in the block between clusters $k$ and $l$, i.e. $p_{kl}=n_k n_l$,
and $y_{kl}$ is the number of directed edges from nodes in cluster $k$ to nodes in cluster $l$.
We also use the Beta function $\text{B}(a,b)=\frac{\Gamma(a)\Gamma(b))}{\Gamma(a+b)}$.

\begin{equation}
\begin{split} %{rl} \displaystyle
\mathrm{P}_\text{SBM}(x,z,K) & = \mathrm{P}_\text{SBM}(K) \times \mathrm{P}_\text{SBM}(z|K) \times \mathrm{P}_\text{SBM}(x|z,K) \\
& = \frac1{K!} \frac1e \times \frac{ \Gamma(K\alpha) }{ \Gamma( N + K\alpha ) } \prod_{k=1}^K \frac{ \Gamma(n_k+\alpha) }{ \Gamma( \alpha ) }
\times
% \mathrm{P}_\text{SBM}(x|z,K)
\prod_{k=1}^K
\prod_{l=1}^K
\frac{\text{B}(y_{kl}+\beta_1,p_{kl}-y_{kl}+\beta_2)}{\text{B}(\beta_1,\beta_2)}
\end{split}
\label{EQpSBM}
\end{equation}

where $\alpha$ is the user-specified parameter to the Dirichlet prior (eq. \ref{EQdirichlet}). %; we use $\alpha=1$ in our experiments.

% The SBM expression for (x,z,K) can be written down and calculated easily in closed form as above.
% But this is not the case with the SCF, as this second factor cannot be solved so easily:
% \begin{equation}
	% f(x,z,K) = \mathrm{P}_\text{SBM}(x,z,K) \times \mathrm{P}_\text{SBM}(v(\pi)=1|x,z,K)
% \label{EQfShort}
% \end{equation}
In a conventional Metropolis-Hasting algorithm (as in \cite{McDaidSBM}), it is convenient to have closed
form expressions of the posterior mass at each state in the chain.
However, it is not necessary to have such expressions and we will see in the next section
how we can work around this and develop a Markov Chain with the correct transition probabilities for the SCF
even though we do not have a fully closed-form expression for $f(x,z,K)$.

%% file: algo.tex
\section{MCMC algorithm}
\label{SECalgo}
% [Give some details of the moves \ldots]
In this section, we will describe the algorithm we have used to sample from
the space of $(z,K)$, with probability proportional to $f(x,z,K)$ (Equation \ref{EQfFull}).
We have extended the software we developed in \cite{McDaidSBM} and we direct the reader
to that paper for detailed definition of all the moves.
% We will summarize the relevant details here where appropriate.

% The Markov Chain visits the various states $(z,K)$, beginning at an arbitrary initial state
% and performing a walk around those states. The goal is to set up the walk such that
% the stationary distribution of the chain is proportional to $f(x,z,K)$.

\subsection{Algorithm for the SBM}

We will first summarize the procedure used in our SBM algorithm, and then describe
the change necessary to turn it into an SCF algorithm.
This means our initial goal is to describe an algorithm whose stationary distribution is
proportional to $\mathrm{P}_\text{SBM}(x,z,K)$.
We define a \emph{proposal distribution} which, given a current state $s=(z,K)$,
will propose a new state $t=(z',K')$.
% We then define \emph{acceptance probability} as a function of the proposal
% probability and the desired stationary distribution to accept some proposals
% and reject others.

The proposals are defined by $p$, where $p_{st}$ is the probability that, given
the chain is in state $s=(z,K)$, that it will propose to move to state $t=(z',K')$.
Clearly, $\sum_t p_{st} \; = \; 1$ for all $s$.
% We are free to define any proposal function we wish, but some perform faster than others.
Given that a proposal has been made to move from $s$ to $t$, where $s \neq t$,
we define an \emph{acceptance probabality} $a_{st}$.
When the proposal is made, we will decide whether to accept or reject the proposal
using a Bernoulli variable with probability $a_{st}$.

% Our desired stationary distribution is $f$, and we can achieve this by defining our
% acceptance probabilities such that \emph{detailed balance} is satisfied:
%
% \begin{equation*}
	% \frac{f(t)}{f(s)} = \frac{p_{st}}{p_{ts}} \frac{a_{st}}{a_{ts}}
% \end{equation*}

In the SBM, where the desired stationary distribution is proportional to $\mathrm{P}_\text{SBM}(x,z,K)$,
we were able to use a standard Metropolis-Hasting \cite{HastingsMetropolis} algorithm with acceptance probability

\begin{equation}
	a_{st} = \min \left( 1, \frac{p_{ts}}{p_{st}} \frac{\mathrm{P}_\text{SBM}(x,t)}{\mathrm{P}_\text{SBM}(x,s)} \right)
\label{EQacptSBM}
\end{equation}
where $\mathrm{P}_\text{SBM}(x,s)$ is defined as $ \mathrm{P}_\text{SBM}(x,z,K)$
and $\mathrm{P}_\text{SBM}(x,t)$ is defined as $ \mathrm{P}_\text{SBM}(x,z',K')$.
%This was possible because we have a closed-form expression for $ \mathrm{P}_\text{SBM}(t) $ (Equation \ref{EQpSBM}).
For the SBM, the transition probabilities satisfy \emph{detailed balance}:
\begin{equation}
	\frac{ t^\text{SBM}_{st} }{ t^\text{SBM}_{ts} } = \frac{ p_{st} a_{st} }{ p_{ts} a_{ts} } = \frac{ \mathrm{P}_\text{SBM}(K,(z',K')=t) }{ \mathrm{P}_\text{SBM}(K,(z,K)=s) }
	\label{EQsbmDetailedBalance}
\end{equation}

One of the moves is a simple Gibbs update on the position of one node, $z_i$. Node $i$ is considered
for inclusion in each of the $K$ clusters.
Another move is called M3, which involves proposing a reassignment of all the nodes in
two randomly-selected clusters.
AE is a move which proposes to split a cluster into two, increasing $K$, or merging
two clusters into one, decreasing $K$.
Together, these moves can visit all states $(z,K)$.
For full details see our earlier work \cite{McDaidSBM}, which was based on existing
algorithms \cite{NobileAllocationSampler,WyseFriel}.

\subsection{Algorithm for the SCF}

But our goal is to develop an algorithm for the SCF. We use the following scheme:
First, make a proposal such as those used in the collapsed SBM algorithm \cite{McDaidSBM}.
Second, calculate the `SBM-acceptance probability' according to Equation \ref{EQacptSBM}.
Third, make a draw from a Bernoulli with this probability to decide whether to Reject or to (provisionally) Accept.
If the proposal was rejected, then there is no further work to be done, the proposal has been rejected.
\emph{But}, if the SBM-acceptance probability led to a (provisional) `acceptance', then there is one final step required to decide on rejection or acceptance of the move;
we draw from the posterior of $\pi|x,z',K'$, drawing a new $\pi$ conditioning on the (proposed) new values of $z'$ and $K'$ in state $t$;
we fully accept the new state if and only if the $\pi$ satisfies the SCF validity constraint in Equation~\ref{EQv}.
This procedure is giving in pseudocode in Table~\ref{TBLpseudocode}.

\begin{table}[h]
\center
\begin{tabular}{|l|}
	\hline
\texttt{~~~~~Given current state, $s = (z,K)$ } \\
\texttt{~~~~~Propose new state, $t = (z',K')$ } \\
\texttt{~~~~~Calculate SBM-acceptance propability, $a_{st}$ } \\
\texttt{~~~~~Draw a Bernoulli with probability $a_{st}$. } \\
\texttt{~~~~~If Failure: } \\
\texttt{~~~~~~~~~REJECT } \\
\texttt{~~~~~Else: } \\
\texttt{~~~~~~~~~Draw $\pi|x,z',K'$ from posterior } \\
\texttt{~~~~~~~~~Test if $\pi$ satisfies $v(\pi$) } \\
\texttt{~~~~~~~~~If Satisfactory: } \\
\texttt{~~~~~~~~~~~~~ACCEPT } \\
\texttt{~~~~~~~~~Else: } \\
\texttt{~~~~~~~~~~~~~REJECT } \\
	\hline
\end{tabular}
\caption{Pseudocode describing the acceptance and rejection rules in the SCF algorithm}
\label{TBLpseudocode}
\end{table}

In this algorithm, a proposal $s \rightarrow t$ (with $s \neq t$) will only be
accepted if the SBM-acceptance succeeds \emph{and} if the $\pi|x,z,K$ satisfies the constraint.
Given that the current state is $s$, the probability of transitioning to another state $t$
is
$$ t^\text{SCF}_{st} = p_{st} \times a_{st} \times  \mathrm{P}_\text{SBM}(v(\pi)=1|x,z',K') $$

We will shortly show that this algorithm is correct for drawing from the desired stationary
distribution, but first we describe how to draw $\pi$ from the its posterior given $(x,z,K)$.
$\pi$ is a $K \times K$ matrix, one element for each block.
In this posterior, as in the prior, these elements are independent of each other and
therefore we proceed by estimating each element of $\pi$ separately.
% we need only to consider the prior and the observed data which is relevant for that element.
The prior on each element of $\pi$ is, as described earlier, a Beta($\beta_1,\beta_2$).
The data for that block is the number of edges which appears, $y_{kl}$, and
the number of non-edges that are in that block, $p_{kl}-y_{kl}$.
In this case, the posterior is Beta($\beta_1+y_{kl},\beta_2+p_{kl}-y_{kl}$).
For each element in $\pi$, this posterior Beta is prepared and one draw is made from each.
If the elements on the diagonal, $\pi_{mm} \sim \text{Beta}(\beta_1+y_{mm},\beta_2+p_{mm}-y_{mm})$,
are greater than those off the diagonal, $\pi_{kl} \sim \text{Beta}(\beta_1+y_{kl},\beta_2+p_{kl}-y_{kl})$,
then the move is accepted.

Now, we show that this satisfies detailed balance and that the stationary distribution
is proportional to $f(x,z,K)$.
% To do this, we show that the ratio of the transition
% probabilities between two states is the same as the ratio of the stationary distribution
% at the two states.
We reuse Equation \ref{EQsbmDetailedBalance} in this proof:

\begin{align*}
	\frac{t^\text{SCF}_{st}}{t^\text{SCF}_{ts}}	= & \frac{ p_{st} \times a_{st} \times  \mathrm{P}_\text{SBM}(v(\pi)=1|x,(z',K')=t) }{ p_{ts} \times a_{ts} \times  \mathrm{P}_\text{SBM}(v(\pi)=1|x,(z,K)=s) } \\
	% =  &
	% \frac{ p_{st} \times a_{st} }{ p_{ts} \times a_{ts} } \times \frac{ \mathrm{P}_\text{SBM}(v(\pi)=1|x,(z,K)=t) }{ \mathrm{P}_\text{SBM}(v(\pi)=1|x,(z,K)=s) } \\
	=  &
	\frac{ \mathrm{P}_\text{SBM}(x,(z',K')=t) }{ \mathrm{P}_\text{SBM}(x,(z,K)=s) } \times \frac{ \mathrm{P}_\text{SBM}(v(\pi)=1|x,(z',K')=t) }{ \mathrm{P}_\text{SBM}(v(\pi)=1|x,(z,K)=s) } \\
	=  &
	\frac{ f(x,(z',K')=t) }{ f(x,(z,K)=s) }
\end{align*}

% This shows that this algorithm satisfies detailed balance and therefore draws from the posterior,
% given the network $x$, of the clustering $z$ and the number of clusters $K$.
% In conventiontal Metropolis-Hasting, the acceptance probability can be written down
% and calculated exactly in software. In the method presented here, we are not able
% to write down such an expression, but nonetheless we have developed an algorithm
% with the correct transition probabilities.

We also use a method of label-switching
which was introduced in \cite{NobileAllocationSampler} and which we used in \cite{McDaidSBM}.
The chain will often visit states which are essentially equivalent to earlier states,
but where the cluster labels have merely been permuted.
The procedure involves permuting the labels of the clusters with the goal of maximizing
the similarity of the latest state to all the previous states.
This leads to more easily interpretable results from the chain.

If it is possible to solve Equation \ref{EQfFull} exactly, this would probably allow
us to have larger acceptance probabilities and to increase the speed of the algorithm
accordingly.
Currently, the algorithm can, in theory, get trapped for some time in a state where
the constraint typically fails for that state and for neighbouring states, making
it difficult for the algorithm to climb towards better states.
This is worth some further consideration, and perhaps an algorithm based on an uncollapsed representation might be best.
A naive uncollapsed algorithm, where just one of $z$ or $\pi$ or $\theta$ is updated in a move, would mix very slowly.
It may be possible to use moves such as those in the allocation sampler to propose
changes simultaneously to the clustering $z$ and to the density matrix $\pi$ and to the cluster-membership-probability vector $\theta$;
such an algorithm may mix as well as the allocation sampler;
such a method would also make it easier to efficiently handle the constraint.
However, this method would be complex to implement; it may be worthwhile to investigate this further.

%% file: synth.tex
\section{Evaluation with synthetic data}
\label{SECsynth}
\begin{figure}[b]
	\centering
	\includegraphics[trim=4.0cm 4.0cm 4.0cm 4.0cm, clip=true, width=0.6\columnwidth]{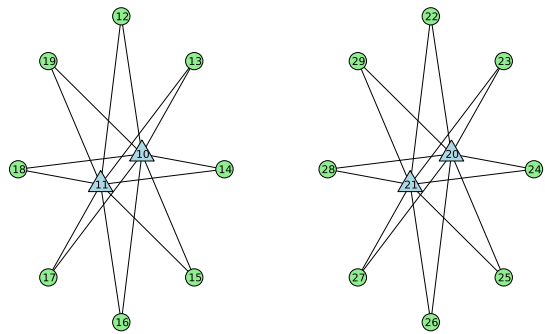}
	\caption{The `$2\times2$' network. Two `roles', peripheral and central. And two communities also, left and right. The SCF finds the two communities, and the SBM finds the roles. }
	\label{FIGlayout2x2}
\end{figure}

% In this section, we evaluate the SCF on two simple synthetic networks.
In this section, we evaluate the SCF on a simple synthetic network.
We compare the results with those found by the basic SBM algorithm.
If we generate data strictly according to the generative SCF model, then both
algorithms tend to be quite accurate, see our earlier work \cite{McDaidSBM} for a detailed
analysis of the accuracy of the collapsed SBM MCMC algorithm.
Therefore, in order to challenge the algorithms, instead we
construct a network where the SBM and SCF get different results
in order to demonstrate the preference of the SCF for `community-like' structure.
We consider the undirected network in Figure~\ref{FIGlayout2x2}, which has
two star-like communities.
Each of these communities has ten nodes, made up of two central
nodes and eight peripheral nodes.
Every central node is connected to every periphery node.

This network has a more heterogenous degree distribution;
this very loosely approximates the heavy-tailed degree distribution
seen in many real-world networks.
If we generate data strictly according to the SBM or SCF
the degree distribution is more homogenous, especially the
distribution of the degrees within a single cluster.
% \begin{figure}
	% \centering
	% \includegraphics[width=0.8\columnwidth]{FIGlayout3x3}
	% \caption{The `3x3' network. Three `roles'; 16 peripheral nodes, 4 mid-level nodes and 1 central node in each of the three communities.}
	% \label{FIGlayout3x3}
% \end{figure}

In all the experiments in this section and the following section, we ran the algorithm for 10,000,000 iterations.
By default, we allow the algorithm to select
the number of clusters itself as the allocation~sampler
algorithm naturally searches the entire search space.
With this network, the SCF selects $K=2$ and it clusters the nodes into the
two star-like communities.
The Markov Chain spends 97.5\% of its iterations in that `ground truth' state.

\begin{figure}
	\centering
	\includegraphics[width=0.32\columnwidth]{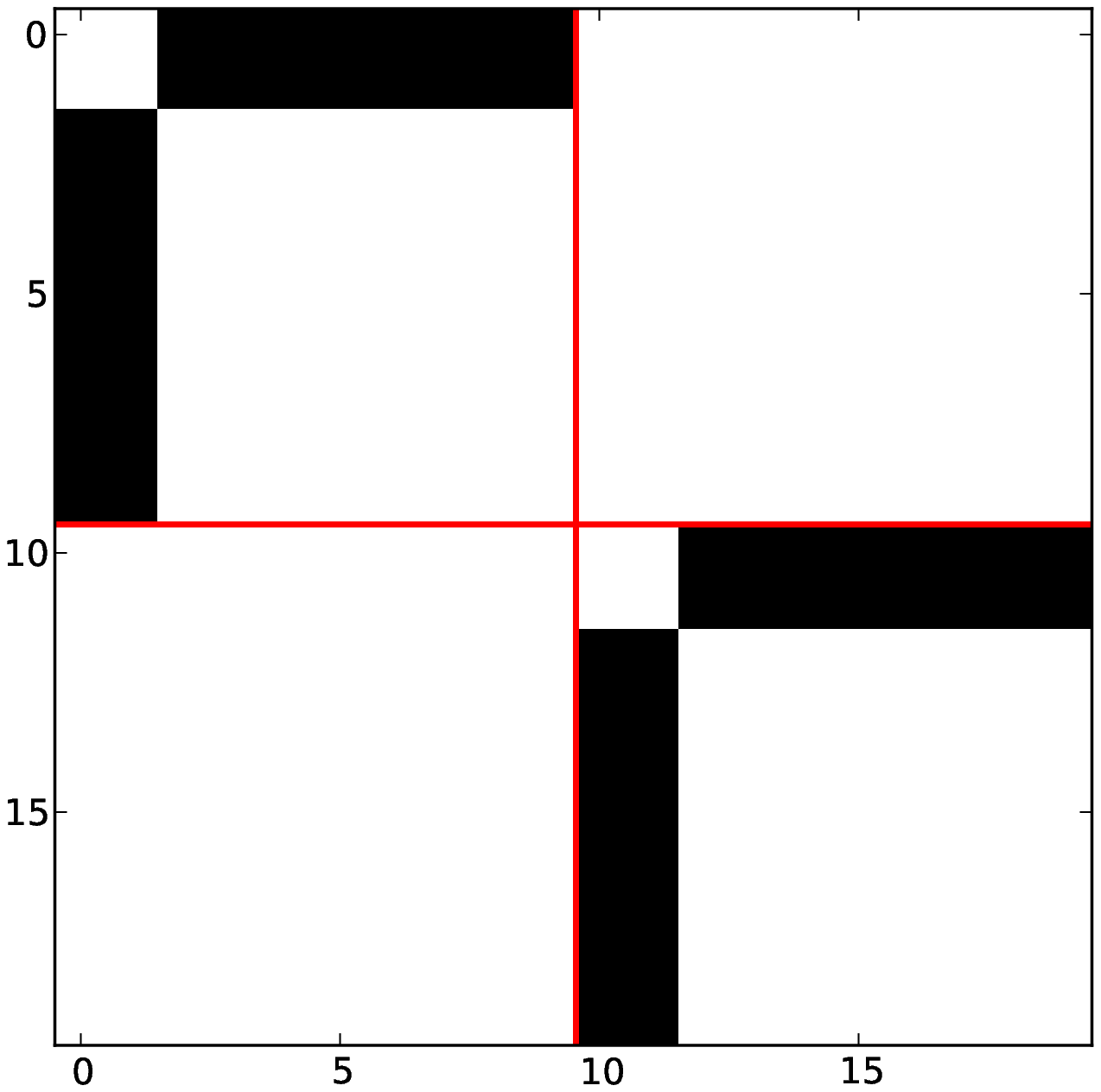}
	\includegraphics[width=0.32\columnwidth]{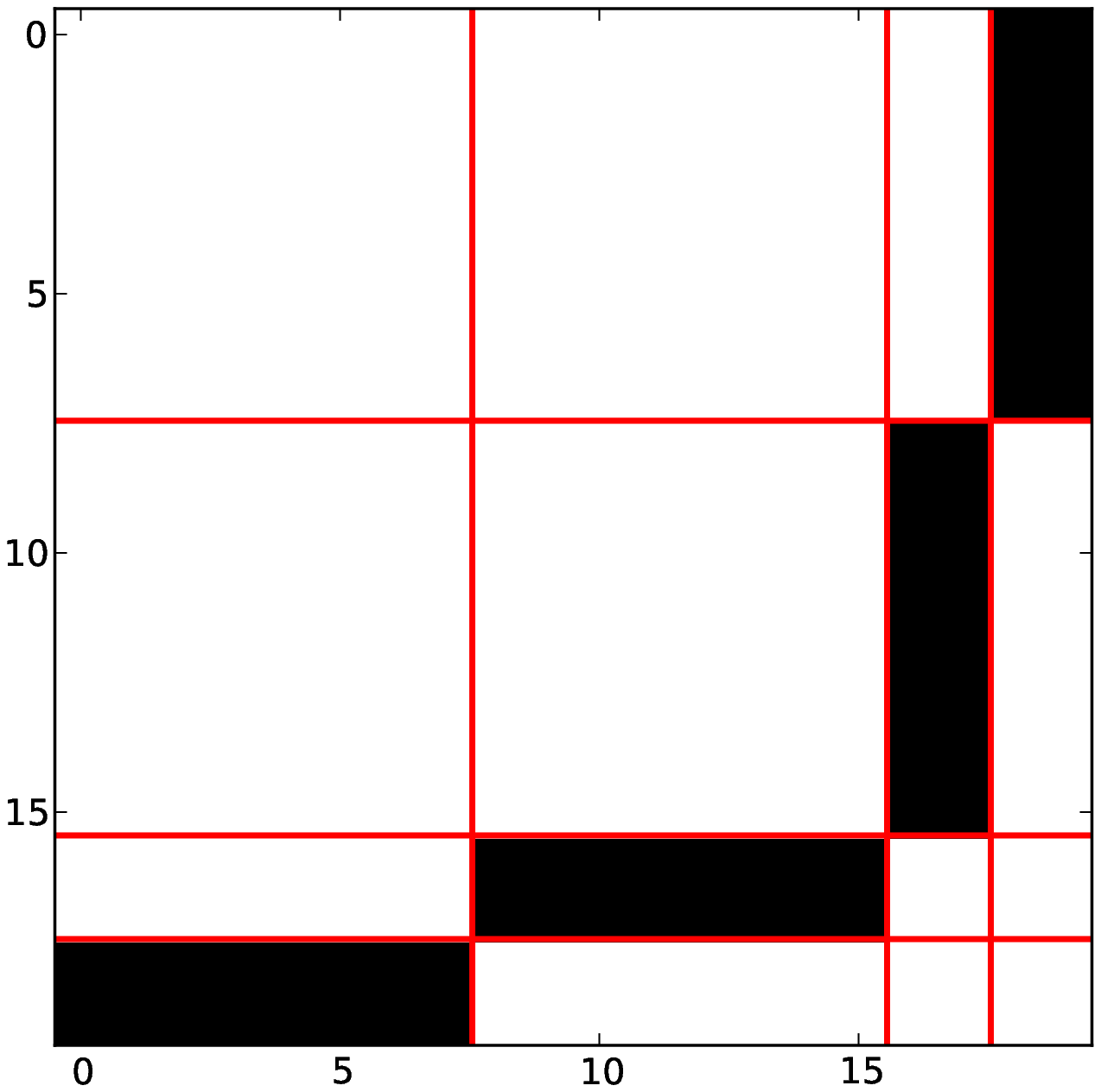}
	\includegraphics[width=0.32\columnwidth]{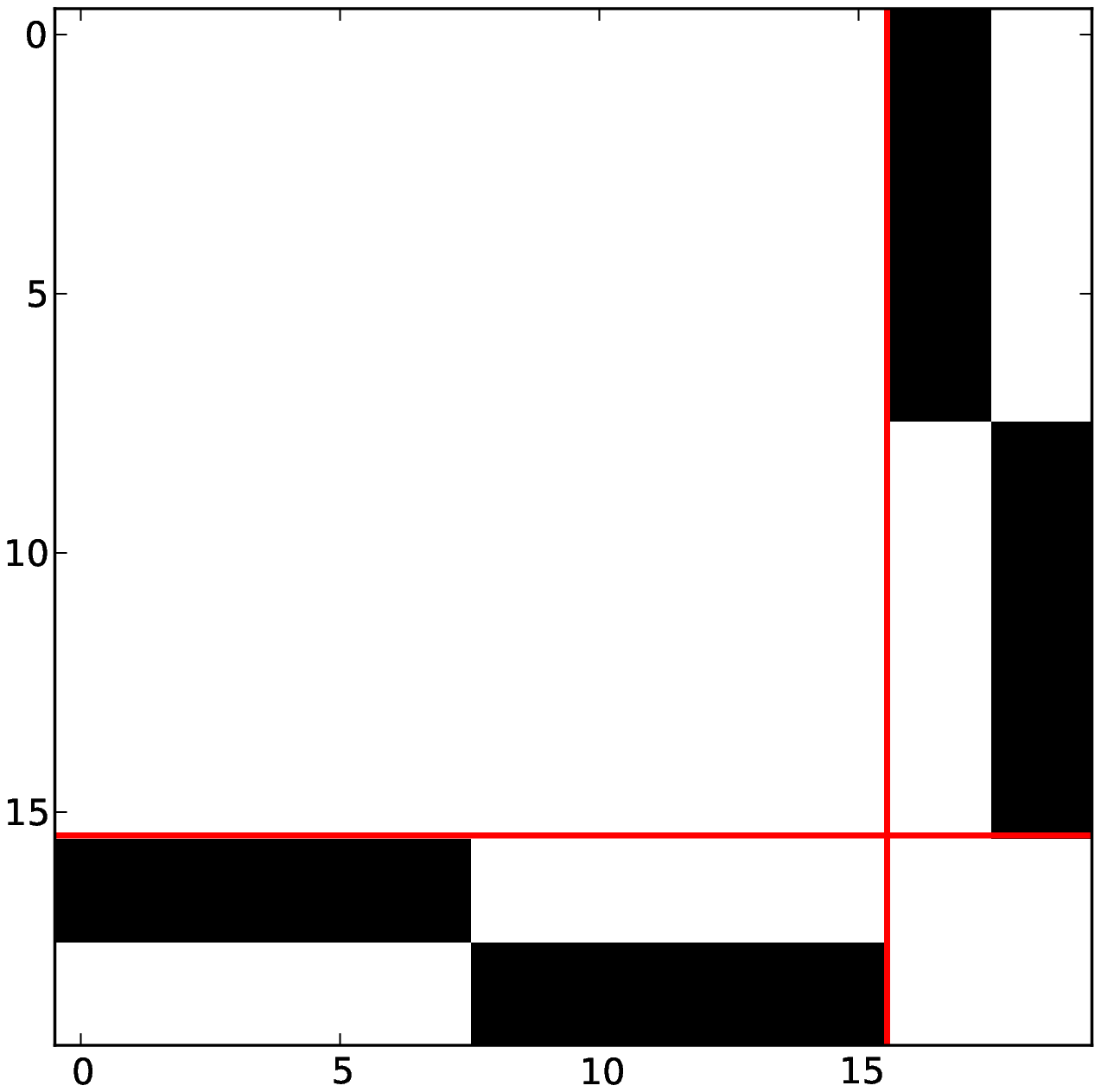}
	\caption{The adjacency matrices showing the clusterings found by the SCF (left) and SBM (middle) on the `$2\times2$' network (Figure~\ref{FIGlayout2x2}). The SCF has found the communities, with all edges inside the clusters, as expected. The SBM has divided the nodes according to degree and community, but there are no edges within any of the four clusters found by the SBM. On the right, we see how the SBM finds only the roles if the number of clusters is fixed at $K=2$ in advance. Only the SCF has placed all the edges inside the clusters and correctly estimated the number of communities.}
	\label{FIG2x2adjSBMSCFk0}
\end{figure}

On the other hand, the SBM select 4 clusters. %, see Figure~\ref{FIG2x2adjSBMSCFk0}.
It subdivides each of the two true communities into
two further communities - one containing the central nodes
and the other containing the peripheral nodes.
We see this in Figure~\ref{FIG2x2adjSBMSCFk0}, where
very few edges are inside the found clusters.
Even if we restrict the SBM to consider only $K=2$, then it again divides the nodes into central and periphery nodes.
Regardless of the number of clusters, the SBM finds clusters which do not contain any of the edges;
this is the opposite of what we expect in community finding.

In networks there may be multiple types of structure that can be detected; the SCF focuses on finding
the `community-like' structure, where the clusters are expected to be internally dense.
In synthetic and empirical networks with a heavy-tailed degree distribution
the SBM may have a tendency to cluster nodes according to their
degree, or other structural roles, and not according to community structure.

%% file: karate.tex
\section{Empirical networks}
\label{SECkarate}
In this section, we apply the SCF to two well-known social networks.

\subsection{Sampson's Monks}
Sampson \cite{SampsonsMonks} gathered data on novices at a monastery\footnote{Sampson's monk data as an R package: \url{http://rss.acs.unt.edu/Rdoc/library/LLN/html/Monks.html}}.
There are 18 novices in the network and a pair are linked if they reported a positive friendship between them, giving us an undirected network.
There were factions within the group, which Sampson labelled \emph{Loyal Opposition}, \emph{Young Turks} and \emph{Outcasts}.

\begin{figure}
	\centering
	\includegraphics[trim=5.9cm 1.9cm 3.9cm 1.9cm, clip=true, width=0.45\columnwidth]{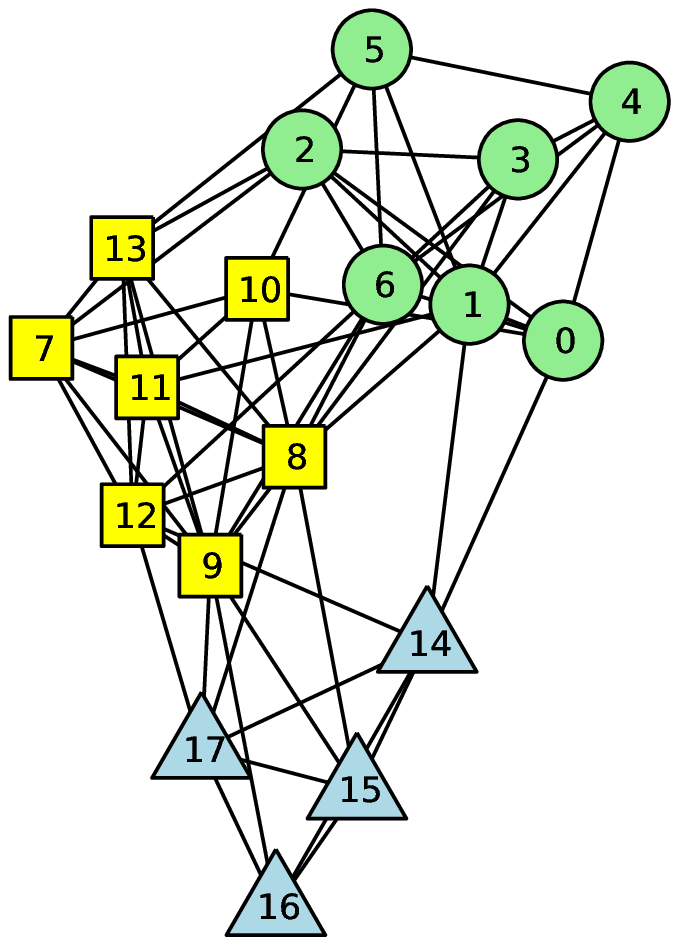}
	\includegraphics[trim=3.0cm 0.0cm 2.0cm 1.5cm, clip=true, width=0.45\columnwidth]{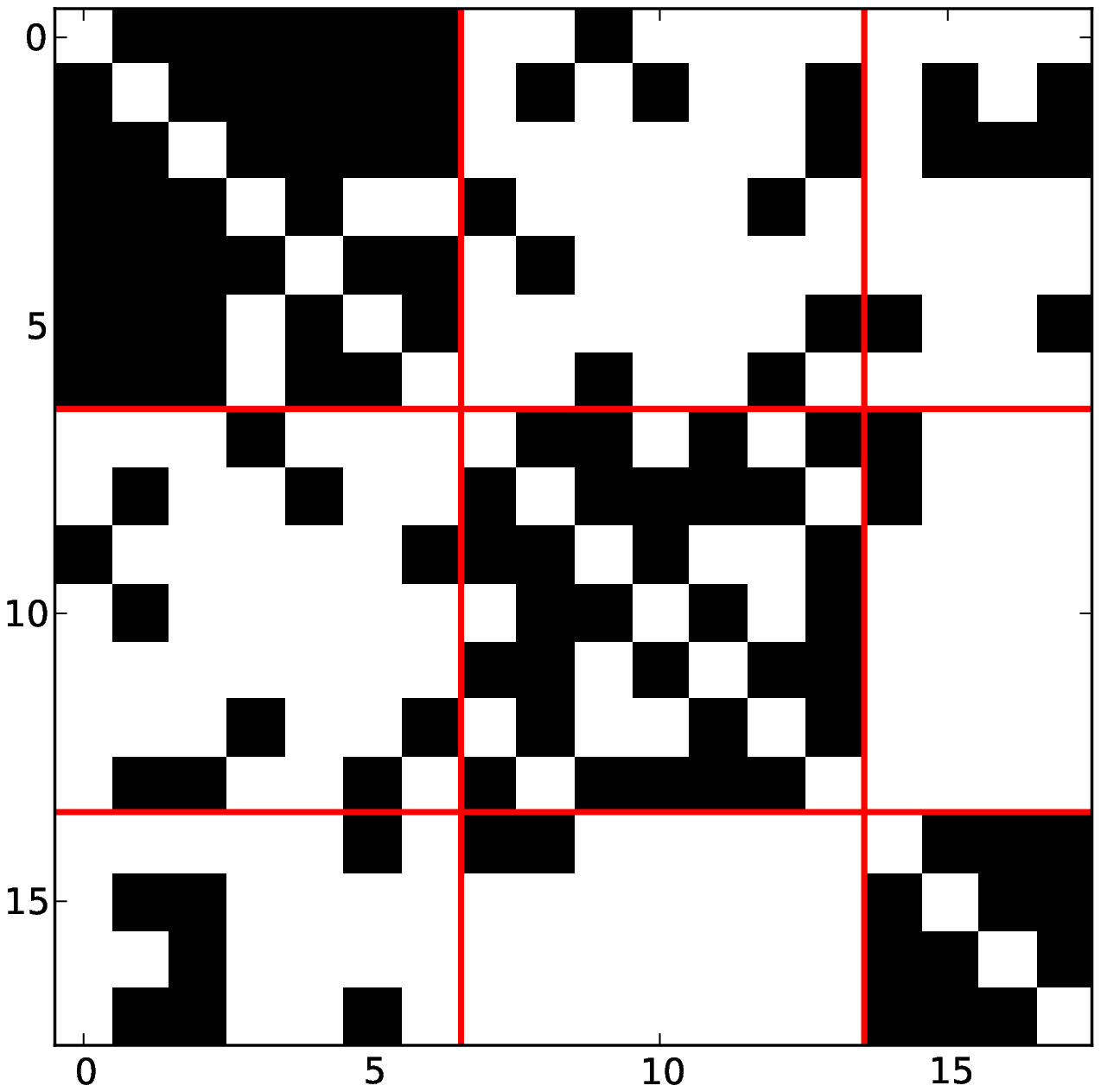}
	\caption{Sampson's monks, and the 3-way split found by the SCF which matches with the factions found by Sampson.}
	                                                                           \label{FIGsampson000K0SCF}
\end{figure}

We ran the SCF method on this dataset for 10,000,000 iterations.
It estimated the number of clusters at 3, with 88.5\% of the iterations.
For 69\% of the iterations, the clustering was exactly equal to the factions reported by Sampson.
The network and adjacency matrix are shown in Figure~\ref{FIGsampson000K0SCF}.
We also ran the data through the SBM. It found very similar results. This suggests that if
the community structure is strong, then either algorithm can detect it. However,
the SBM is slightly less accurate and only 55\% of the iterations involve $K=3$.
This suggests that there is other structure, perhaps the high-degree versus 
low-degree structure, that is trying to assert itself.

\subsection{Zachary's Karate club}
Now, we apply the SCF to a network of interactions at a karate club \cite{ZacharyKarate}, again demonstrating
the ability of the SCF to detect community structure where the SBM focusses on other types of structure.

\begin{figure}[h]
	\centering
	\includegraphics[trim=3.9cm 3.9cm 3.9cm 3.9cm, clip=true, width=0.43\columnwidth]{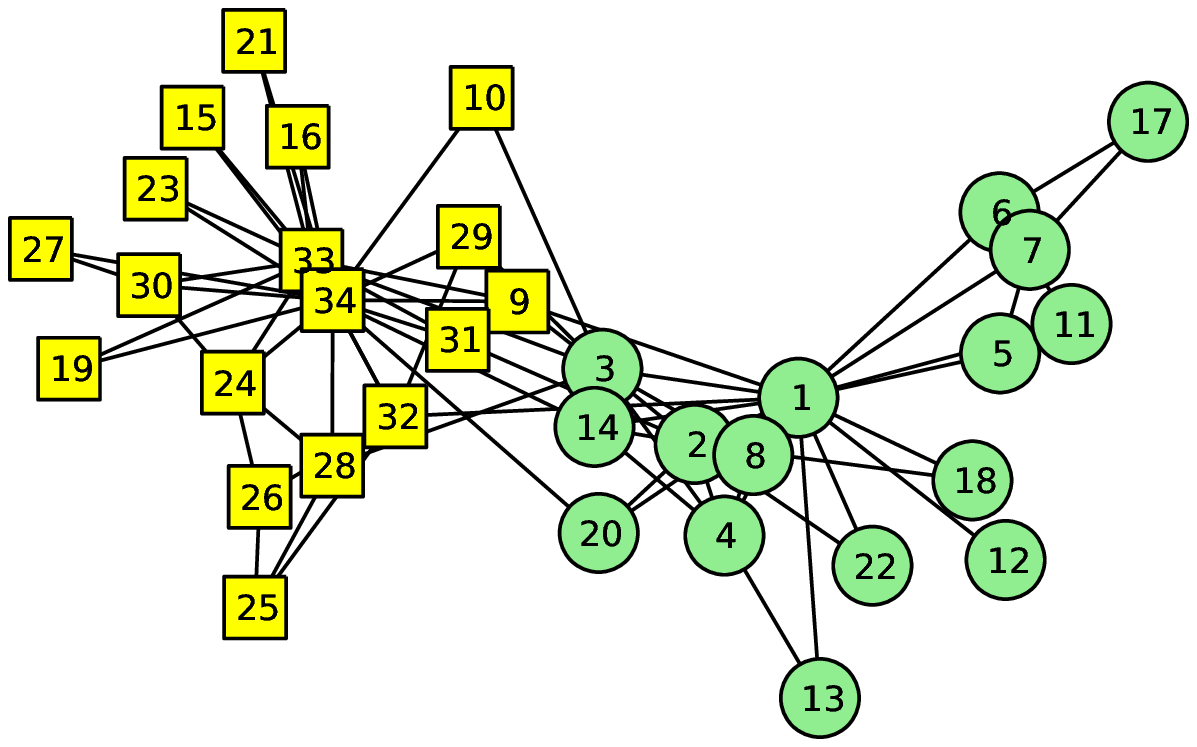}
	\includegraphics[width=0.43\columnwidth                                         ]{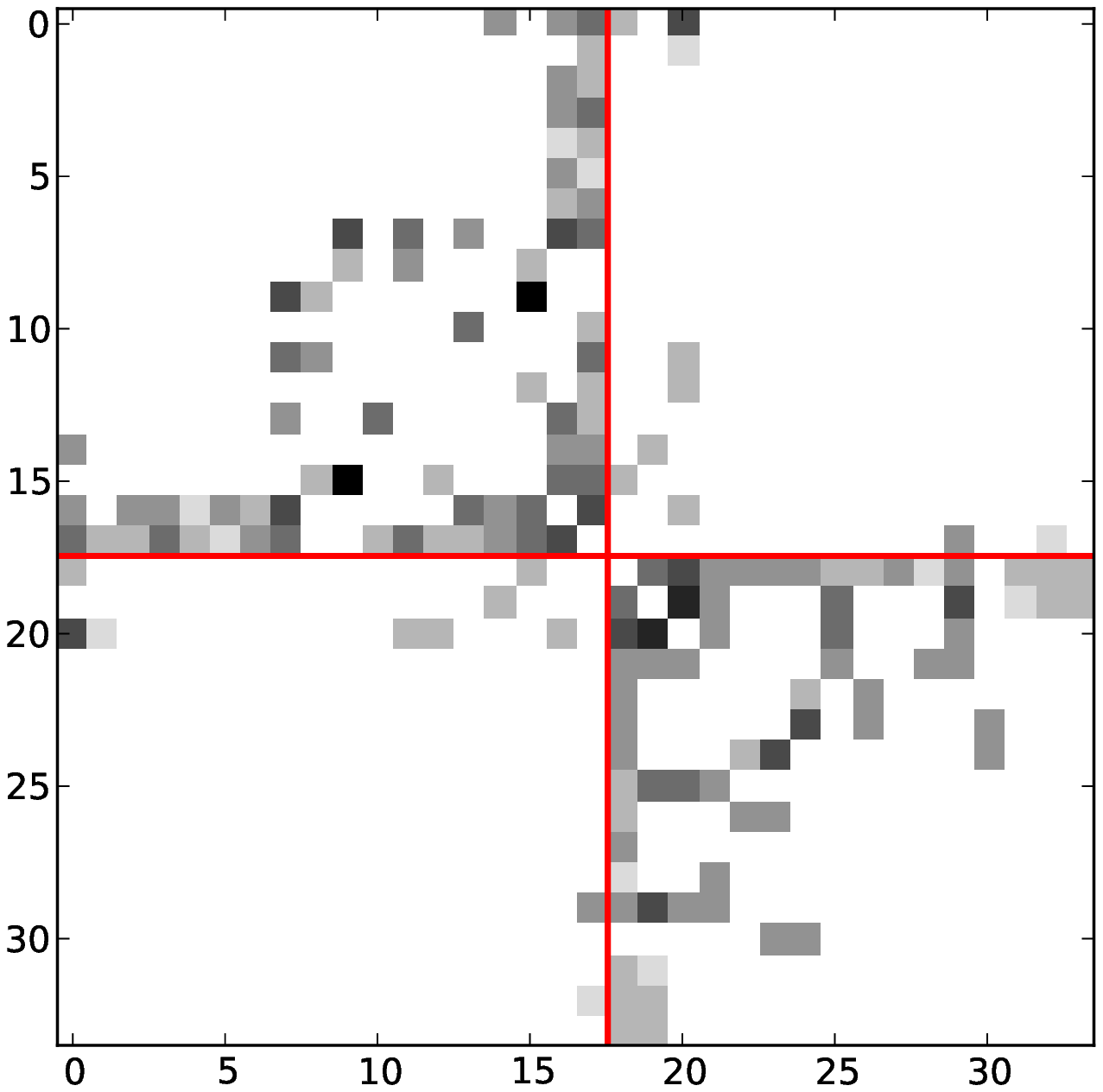}
	\caption{The karate club network of \cite{ZacharyKarate}. The width indicates the strength of the relationship by counting the
	number of distinct interaction types recorded between the two members. The club split in two after the survey and the colour
	of the node records the split.
	On the right is the adjacency matrix of this network. The rows and columns have been ordered according to which faction
	the node is in; most of the edge weight is on the top-left and bottom-right, as would be expected in good community structure.
	The SCF algorithm finds this clustering when $K=2$.
	}
	\label{FIGlayoutKarateGT}
\end{figure}

The members of the karate club were asked about their social interactions with other members,
focusing on interactions outside of the lessons and tournaments.
This gives us a network of 34 members and 78 interactions.
The interaction data is weighted, according to the number of
distinct social interaction types reported by the members; a larger number
is taken to indicate a stronger friendship\footnote{Weighted karate club network: \url{http://vlado.fmf.uni-lj.si/pub/networks/data/Ucinet/zachary.dat}}.
After the survey was taken, the club split into two factions over a dispute of the cost of the lessons.
The network is visualized in Figure~\ref{FIGlayoutKarateGT}.

This network has weighted edges and hence we apply our SCF constraint (eq.~\ref{EQv}) to the
weighted variant of the SBM. The edges have a Poisson weight, and the rate of the Poisson, $\pi_{kl}$, is different from each block and
comes from a Gamma prior; full details of this edge model are in the Appendix of our earlier work\cite{McDaidSBM}.

\begin{figure}
	\centering
	\includegraphics[trim=3.9cm 3.9cm 3.9cm 3.8cm, clip=true, width=0.43\columnwidth]{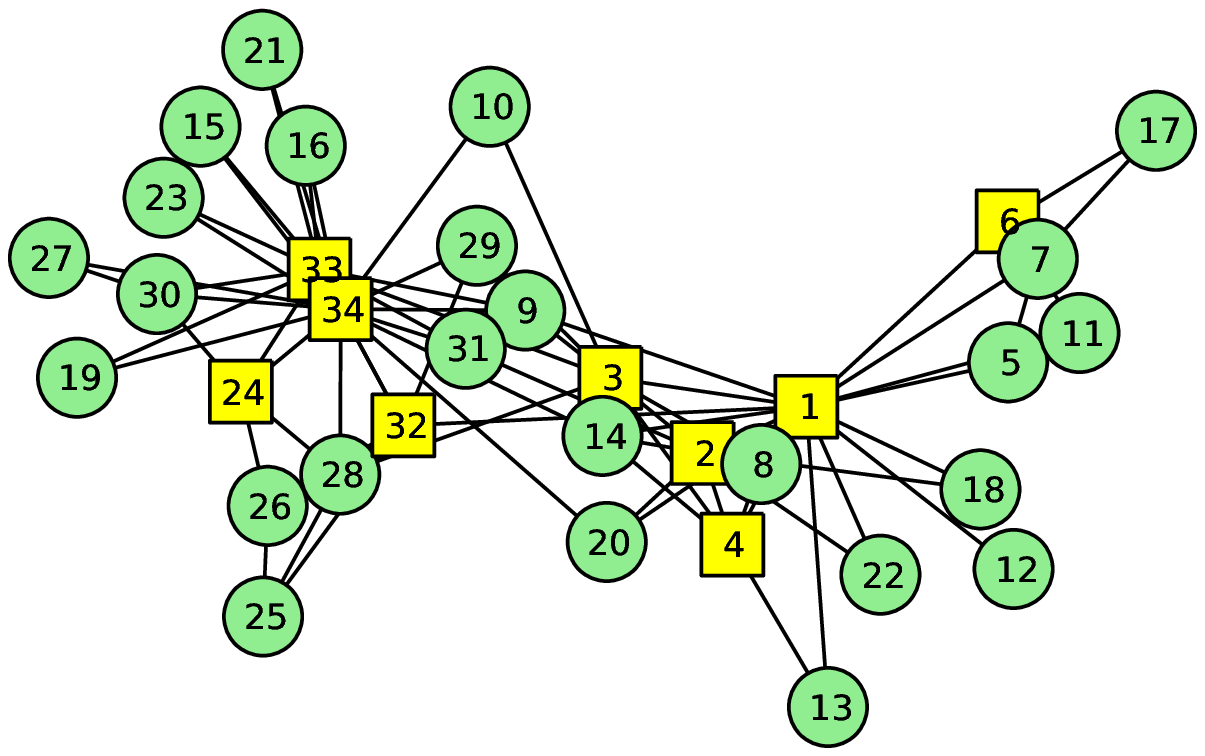}
	\includegraphics[width=0.43\columnwidth                                         ]{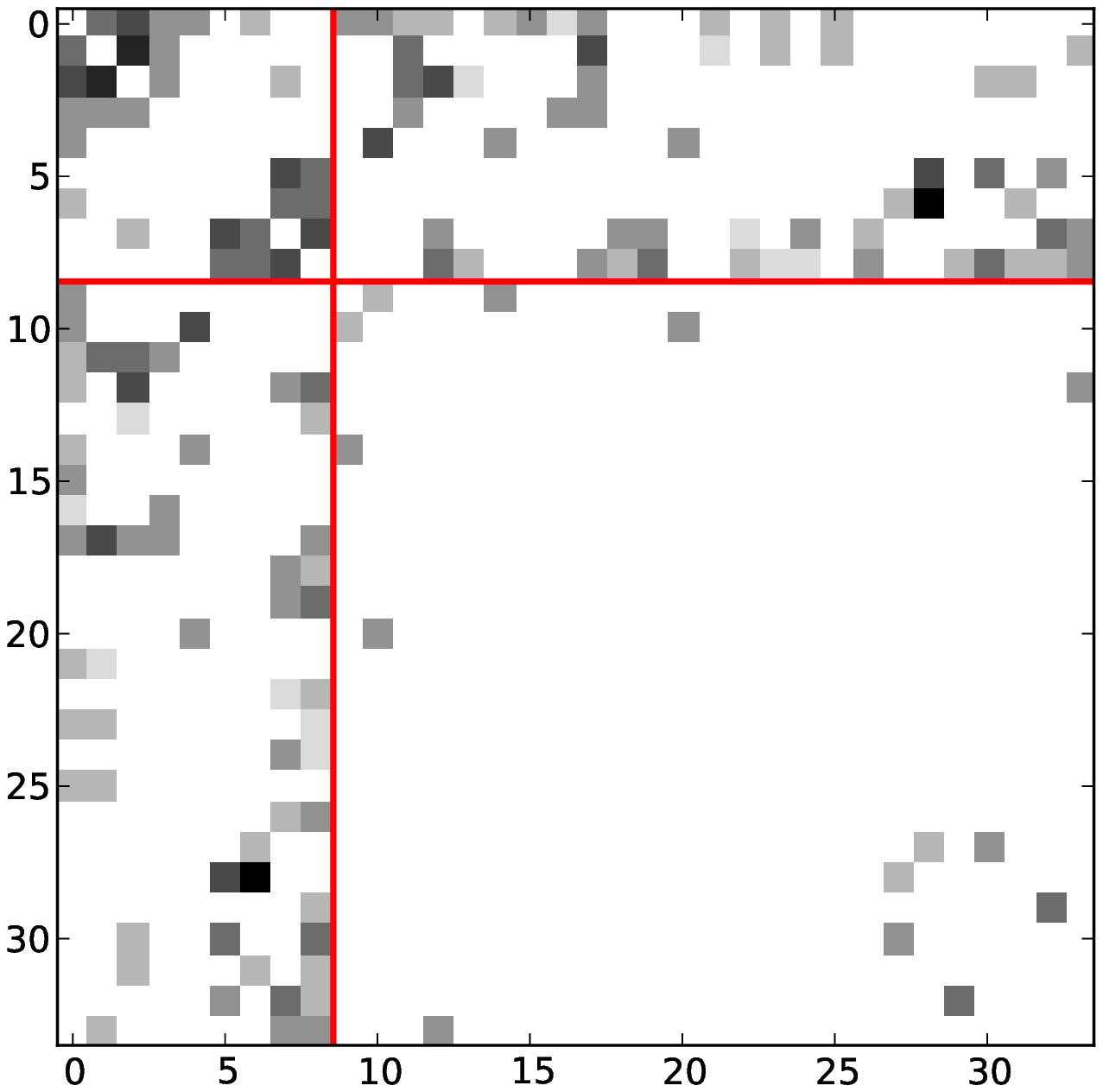}
	\caption{The SBM, when told to find two clusters, divides the nodes according to degree.}
	                                                                           \label{FIGkarate000K2SBMw1}
\end{figure}

If we fix the number of clusters at $K=2$,
then the SCF will correctly cluster the nodes according the split that occured in the club;
the chain will spend 85.5\% of its iterations in that state.
This contrasts with the SBM, which instead clusters the nodes into 9 high-degree and 25 low-degree
nodes, a clustering which is quite different from the factional split; this SBM clustering is in Figure~\ref{FIGkarate000K2SBMw1}.
The high-degree nodes include the leaders of each faction.

Unfortunately, unlike our earlier networks, the SCF does not correctly estimate the number of clusters within
the karate club. We had to specify that $K=2$ in order to find the correct clustering, whereas our MCMC algorithm estimates $K=5$.
The issue of model selection within this model may be worth considering further.

%% file: conclusion.tex
\section{Conclusion}
\label{SECconclusion}

Community finding is popular in the social science literature, but many statistical models are defined for
block-modelling, not explicitly for community-finding.
In order to investigate community-finding, we have introduced a constraint that the density inside clusters be larger than the density
between pairs of clusters.
We have extended an existing block-modelling method, which was based on the Stochastic Block Model (SBM),
to take account of this constraint.
We evaluated the method and shown it can detect community structure where the SBM cannot.

%There may be multiple different types of structure in a network, and it may be desirable
%to specify which type of structure you wish to detect. We have demonstrated two types:
%clustering-by-degree and clustering-by-community-finding.
%Other constraints may be considered also, perhaps the user may wish to enforce a different
%ordering on the values in the $\pi$ matrix to the ordering we have specified.

% There are many variants of the SBM, and the SCF constraint may be applied to them.
% We have considered inter-weighted and unweighted networks.
% We have focussed here on the simplest variant of the SBM, where edges are unweighted and undirected.
% The collapsing techniques employed here, and the algorithmic techniques, can be applied to other variants.